\documentclass{elsart}
\usepackage{epsfig}

\begin{document}
\begin{frontmatter}
\title{Random Walk Beyond Hartree-Fock}
\author{J. E. Gubernatis and M. Guerrero\thanksref{DOE}}
\address{Theoretical Division, 
Los Alamos National Laboratory,
Los Alamos, NM 87545}
\thanks[DOE]{Both supported by the U. S. Department of Energy.}

\begin{abstract}
We give a brief discussion of the recently developed Constrained-Path
Monte Carlo Method. This method is a quantum Monte Carlo technique
that eliminates the fermion sign problem plaguing simulations of
systems of interacting electrons. The elimination is accomplished by
trading an exact procedure for an approximate one that has been
demonstrated to give very accurate estimates of energies and many-body
correlation functions. We also give a short review of its
applications, a discussion of several strategies for parallelizing it,
and some speculation of its future extensions.
\end{abstract}
\end{frontmatter}

\section{Introduction}
We will give a brief discussion of the recently developed
Constrained-Path Monte Carlo (CPMC) Method \cite{zhang1,zhang2}. This
ground-state ($T=0$) method is a quantum Monte Carlo (QMC) technique
that eliminates the infamous fermion sign problem plaguing simulations
of systems of interacting electrons. The fermion sign problem causes
the variance of measured quantities to increase exponentially with
increasing system size and decreasing temperature. Rapidly the sign
problem destroys one's ability to compute with acceptable accuracy.
With the CPMC method one has simulated system sizes not possible with
the standard method \cite{loh}.  In particular, the lattice size
dependence of many-body superconducting pairing correlations functions
were simulated for some of the largest lattice sizes studied to
date.

In CPMC method the elimination of the sign problem is accomplished by
trading an exact procedure for an approximate one that has been
demonstrated to give very accurate estimates of energies and various
many-body correlation functions. The exact procedure determines the
lowest eigenvalue and eigenvector of the Hamiltonian by projecting
them from a trial state.  This procedure is easily converted to a
branched random walk. Because of the sign problem, the random walkers
carry a positive and negative weight, unfortunately in proportions
such that the average weight (sign) becomes zero as the system size
increases. The constrained path method is a particular way to break
the symmetry in the sign of the walkers and produce only positively
weighted walkers by eliminating those with a negative overlap with a
certain constraining state. The procedure bears a similarly to the
fixed-node Monte Carlo method \cite{reynolds} that has been
successfully used for several decades in simulations of interacting
electrons systems defined in the continuum of configuration space. The
CPMC method however operates in the manifold of single-particle states (Slater
determinants) defined in Fock space and hence represents quite a
different and novel Monte Carlo algorithm.

In the next section we will give a brief discussion of the
method. Then in the following sections will discuss the various models
to which the method has been applied, highlighting significant
results. After this we will discuss several strategies for
parallelizing the method. At first glance it would seem as if we could
simply exploit the natural parallelization enjoyed by most Monte Carlo
methods \cite{bonca2}. In fact we need to do a bit more. In the
closing section we make some speculations on future applications and
extensions of the CPMC method and the potential changes in
parallelization procedures.

\section{The Constrained-Path Monte Carlo Method}

Our numerical method is extensively described and benchmarked
elsewhere \cite{zhang1,zhang2}. Here we only discuss its basic
strategy and approximation.  In the CPMC method, the ground-state wave function
$|\Psi_0\rangle$ is projected from a known initial wave function
$|\Psi_T\rangle$ by a branching random walk in an over-complete space
of Slater determinants $|\phi\rangle$.  In such a space, we can write
$|\Psi_0\rangle = \sum_\phi \chi(\phi) |\phi\rangle$.  The random walk
produces an ensemble of $|\phi\rangle$, called random walkers, which
represent $|\Psi_0\rangle$ in the sense that their distribution is a
Monte Carlo sampling of $\chi(\phi)$, that is, a sampling of the
ground-state wave function.

To completely specify the ground-state wave function for a system of
interacting electrons, only determinants satisfying
$\langle\Psi_0|\phi\rangle>0$ are needed because $|\Psi_0\rangle$
resides in either of two degenerate halves of the Slater determinant
space, separated by a nodal surface ${\bf N}$ that is defined by
$\langle\Psi_0|\phi\rangle=0$.  The degeneracy is a consequence of
both $|\psi_0\rangle$ and $-|\psi_0\rangle$ satisfying Schr\"odinger's
equation. The sign problem occurs because walkers can cross ${\bf N}$
as their orbitals evolve continuously in the random
walk. Asymptotically they populate the two halves equally, leading to
an ensemble that has zero overlap with $|\Psi_0\rangle$.  If ${\bf N}$
were known, we would simply constrain the random walk to one half of
the space and obtain an exact solution of Schr\"odinger's equation.
In the constrained-path QMC method, without {\it a priori\/} knowledge
of ${\bf N}$, we use a trial wave function $|\Psi_T\rangle$ and
require $\langle\Psi_T|\phi\rangle>0$.  This is what is called the
constrained-path approximation.

The quality of the calculation clearly depends on the trial wave
function $|\Psi_T\rangle$. Since the constraint only involves the
overall sign of its overlap with any determinant $|\phi\rangle$, it
seems reasonable to expect the results to show some insensitivity to
$|\Psi_T\rangle$.  Through extensive benchmarking on the Hubbard
model, it has been found that simple choices of this function can give
very good results \cite{zhang1,zhang2}.

Besides as starting point and as a condition constraining a random
walker, we also use $|\Psi_T\rangle$ as an importance function. To
reduce variance, we use $\langle\Psi_T|\phi\rangle$ to bias the random
walk into those parts of Slater determinant space that have a large
overlap with the trial state. For all three uses of $|\Psi_T\rangle$,
it clearly is advantageous to have $|\Psi_T\rangle$ approximate
$|\Psi_0\rangle$ as closely as possible. Only in the constraining of
the path does $|\Psi_T\rangle \not= |\Psi_0\rangle$ generate an
approximation.

Almost all the calculations reported here are done for square lattices
with periodic boundary conditions.  Mostly, we study closed shell
cases, for which the corresponding free-electron wave function is
non-degenerate and is translationally invariant. In these cases, the
free-electron wave function, represented by a single Slater
determinant, is used as the trial wave function $|\psi_T\rangle $.
The use of an unrestricted Hartree-Fock wave function as
$|\psi_T\rangle $ generally produces no significant improvement in the
results.

We remark the the CPMC method has been extended to use generalized
Hartree-Fock wave functions, of which the most famous example is the
BCS wave function. The trick for doing this is described elsewhere
\cite{guerrero1}.

One does simulations because exact solutions are generally
unavailable, and approximate solutions, like those from Hartree-Fock
approximations, are often poor. The objective of the CPMC method and
other simulation methods is to go beyond Hartree Fock
approximation. By expressing the wave function as a linear combination
of Slater determinant, the CPMC method is a type of stochastic
configuration interaction (CI) method. One difference from the classic
CI method is its basis functions, the Slater determinants, being
over-complete. Another difference is the set of basis states is selected via
a constrained, importance-sampled random walk. There is not just one
set of basis functions but many sets. Averaging over these sets is
necessary to compute expectation values.

\section{Applications}

Except for a very recent application to a nuclear physics
model \cite{schmidt}, all applications of the constrained-path method
have been to Hubbard-like models of interest to physical chemists and
condensed-matter physicists. These models can be grouped as one-band,
two-band, and three-band models, and each group usually has targeted
classes of phenomena and materials. The Hubbard models represent
considerable simplifications of the complex interactions
found in actual materials but still represent complicated many-body
problems for which exact solutions are rare and limited to
one-dimensional systems. Under these circumstances Monte Carlo methods
represent perhaps the only controlled means to study the properties of
these models and benchmark approximate theories for these properties.

Hubbard models are used to study a sweeping array of intrinsically
quantum many-electron phenomena. Historically these models were
heavily studied for possible magnetic phenomena (ferromagnetic,
anti-ferromagnetic, paramagnetic, etc.). More recently, they have been
intensively studied for possible representatives of high temperature
superconducting materials. In between, they were studied as candidate
models of heavy fermion and mixed-valence materials.

The principal objects of interest are correlation functions, for
example, spin-spin, charge-charge, pair-pair correlation functions,
because the scaling of these functions with systems size gives a
measure of symmetry of the ground-state. In one and two dimensions,
long-range order at finite temperature is generally precluded by the
Mermin-Wagner theorem. Such states however can exist at zero
temperature, and their existence is signified by the behavior of
correlation functions with increasing systems size. Long range
anti-ferromagnetic order for example would be indicated by the
$(\pi,\pi)$ peak of the Fourier transform of the spin-spin correlation
function increasing with increasing system size and extrapolating to a
non-zero value in the limit of infinite system size. Long-range
superconducting order would be indicated by the long-range part of the
pair-pair correlation function remaining a positive number as the
systems size is extrapolated to infinite volume. Since the sign
problem worsens as the systems size increases, having a good
approximate method, like the CPMC method, is an important tool for
determining whether long-range order exists in these models.

\subsection{One-Band Models}

The classic one-band Hubbard model is given by the Hamiltonian
\begin{equation}
  H =-t\sum_{\langle ij \rangle \sigma} (c_{i \sigma}^\dagger c_{j\sigma}
    + c_{j \sigma}^\dagger c_{i\sigma})
    + U \sum_i n_{i \uparrow} n_{i \downarrow}.
\label{eq:H}
\end{equation}
where the double summation is restrict to nearest neighbors, the
operators $ c_{i \sigma}^\dagger$ and $ c_{i\sigma}$ create and
destroy an electron of spin $\sigma$ at lattice site $i$, and
$n_{i\sigma}=c_{i \sigma}^\dagger c_{i\sigma}$ is the number operator
at site $i$.  Typically the lattices studied have periodic boundary
conditions and were square. When the interaction $U=0$, the model is
exactly solvable and the non-interacting electrons are described by a
single band. When the number of electrons equals the number of lattice
sites, the model is said to be half-filled. At half-filling $\langle n 
\rangle = \langle \sum_\sigma n_{i\sigma}\rangle =1$.

In recent years, this model was extensively studied for its
magnetic and possible superconducting properties. The half-filled
model has no sign problem for most QMC methods, and QMC studies have
played a major role in establishing that the ground state of the
two-dimensional model has long-range anti-ferromagnetic order. This
state is consistent with the observed behavior of the parent (undoped)
state of high temperature superconductors. In these materials
superconductivity appears when the parent state is doped away from
half-filling. At dopings relevant to experiment, the sign problem is
very bad.  With the CPMC method it was possible to study the doped
Hubbard model and to compute various correlation functions as a
function of system size. The main focus has been on superconducting
pairing correlation functions

To be a bit more definite, the types of correlation
functions computed were: The spin density structure factor is
\begin{equation}
S(k_x,k_y) = S({k}) = 1/N\sum_j {\exp}(\mathrm{i}{k}\cdot j)
\langle {s}_{0} {s}_j\rangle,
\label{eq:struct}
\end{equation}
where ${s}_j=n_{j\uparrow}-n_{j\downarrow}$ is the z-component of spin
at site $j$. The charge density structure factor is similar to
(\ref{eq:struct}), with spin replaced by density, i.e., with the $-$
sign in ${s}_j$ replaced by a $+$ sign. The electron pairing
correlation function is defined as
\begin{equation}
P_\alpha (j_x,j_y) = P(j) = \langle \Delta^\dagger_\alpha(j)
\Delta_\alpha({0})\rangle,
\label{eq:pairing_def}
\end{equation}
where $\alpha$ indicates the symmetry of the pairing. The on-site $s$-wave
pairing function has $\Delta_{s}(j) = c_{j\uparrow} c_{j\downarrow}$,
while for $d$-wave pairing we used $\Delta_{d} (j) 
=c_{j\uparrow}\sum_{\delta} f({\delta}) c_{j+
\delta\,\downarrow}$, where ${\delta}$ is $(\pm 1,0)$ and $(0,\pm
1)$. For ${\delta}$ along the $x$-axis, $ f({\delta})$ is $1$;
otherwise it is $-1$.

\begin{figure}[t]
\begin{center}
\epsfig{file=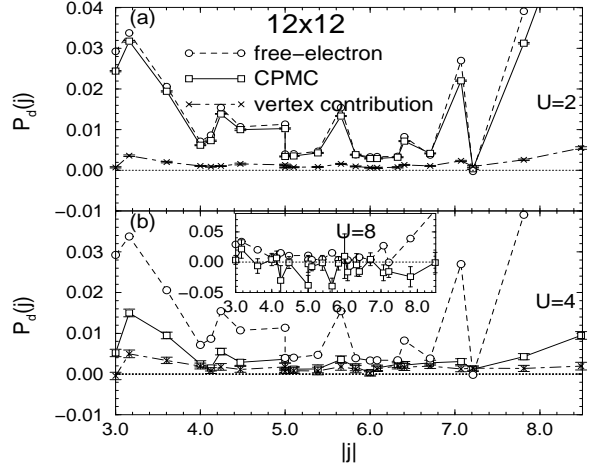,height=2.5in,width=3.0in}
\end{center}
\caption{Long-range behavior of the $d_{x^2-y^2}$ pairing correlation
function versus distance for $\langle n \rangle = 0.85$ and a
$12\times 12$ lattice at $U=2$, 4, and 8 \cite{zhang3}. This behavior
is shown for the free-electron electron and CPMC calculations. Also
shown is the vertex contribution.}
\end{figure}

Simulations of the Hubbard model find that the s-wave pairing
correlations are generally suppressed relative to the d-wave pairing
correlations. For a given lattice size, increasing the value of $U$
suppresses the d-wave pairing and for a given value of $U$, increasing
the lattice size suppresses the d-wave pairing. These results are
inconsistent with long-range order and are illustrated in Figs.~1 and
2 where the d-wave pairing correlation function is shown for a
$12\times 12$ and a $16\times 16$ lattice as a function of distance $|j|$.
In these figures, also reported is the ``vertex contribution'' to the
correlation functions defined as
\begin{equation}
V_\alpha(j) =  P_\alpha(j) - \bar P_\alpha(j)
\end{equation}
where $\bar P_\alpha(j)$ is the contribution of two dressed
non-interacting propagators: For each term in $ P_\alpha(j)$ of the
form $\langle c^\dagger_\uparrow c_\uparrow c^\dagger_\downarrow
c_\downarrow\rangle$, $\bar P_\alpha(j)$ has a term like $\langle
c^\dagger_\uparrow c_\uparrow\rangle\langle c^\dagger_\downarrow
c_\downarrow\rangle$. When combined, Figs.~1 and 2 indicate the likely
absence of long-range superconducting order in the two-dimensional
Hubbard model.

\begin{figure}[b]
\begin{center}
\epsfig{file=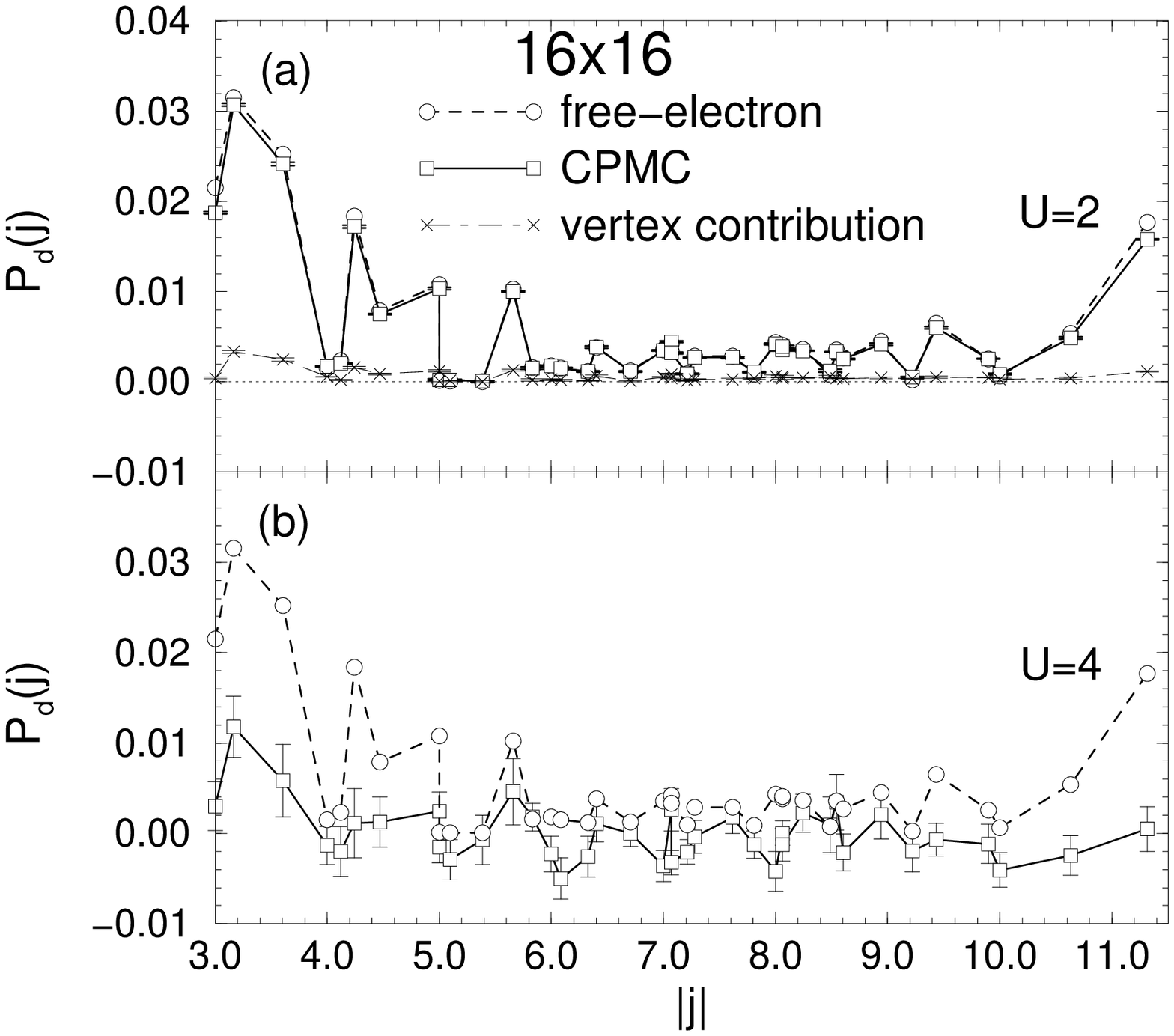,height=2.5in,width=3.0in}
\end{center}
\caption{Long-range behavior of the $d_{x^2-y^2}$ pairing correlation
function versus distance for $\langle n \rangle = 0.85$ and a
$16\times 16$ lattice at $U=2$ and 4 \cite{zhang3}.  This behavior is
shown for the free-electron electron and CPMC calculations. Also shown
is the vertex contribution.}
\end{figure}

Most projector Monte Carlo ground-state calculations of Hubbard models
have projected from trial states that were not superconducting. Using
a trick, we projected from a d-wave BCS superconducting wave function
with a BCS superconducting order parameter $\Delta=0.5$. (Again we use
the trial state not only for the initial state, but also for
importance and constraining states.) Within statistical error, we
found the same d-wave correlation function as we did when we projected
from a free-electron wave function. One of our results is shown in
Fig.~3. This similarity re-enforces the results of Ref.~\cite{zhang3}
that suggest the absence of ODLRO in the two-dimensional Hubbard model. 
\begin{figure}[t]
\begin{center}
\epsfig{file=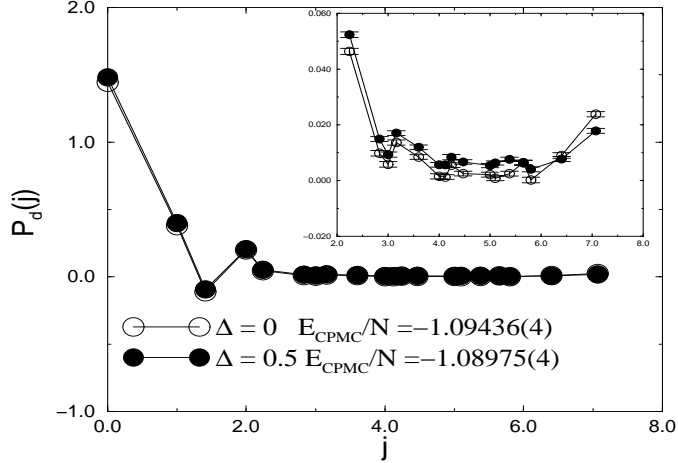,height=2.5in,width=3.5in}
\end{center}
\caption{Pairing correlation functions in the $d_{x^2-y^2}$-wave channel 
given by the CPMC method for a $10\times 10$ lattice, $U=4$, and 
$\langle n \rangle = 0.82$. The inset shows the long range part in detail. The
results are the same for the two different trial wave functions: the
correlations decay quickly with distance \cite{guerrero1}.  On the
other hand the BCS trial wave function exhibits ODLRO. Errors bars are
smaller than the size of the symbols.}
\end{figure}

Adding a next-nearest neighbor hopping term of strength $t'$ to the
classic Hubbard model produces the $tt'U$ model. The CPMC method has
been applied to this model with the results not being materially
different from those without the $t'$. Also adding the $t'$ term,
whose presence is suggested by band structure calculations, does not
seem to change the magnetic properties at half-filling, and away from
half-filling its addition does not seem to enhance superconductivity
\cite{huang,zhang4}.

Adding a nearest neighbor interaction, $V\sum_{\langle ij \rangle}
n_in_j$ with $n_i=\sum_\sigma n_{i\sigma}$, produces the $tUV$
model. The additional interaction generates additional competing
effects that leading to a novel co-existence of states
\cite{mazumdar}.

\subsection{Two-Band Model}

The two-band model is almost always called the periodic Anderson
model. In this model a $d$ and a $f$ orbital are on each lattice
site. These two orbitals per unit cell lead to two bands. The
Hamiltonian is
\begin{eqnarray}
  H &=& -t\sum_{\langle i,j \rangle,\sigma} 
       (d_{i,\sigma}^\dagger d_{j,\sigma}+d_{j,\sigma}^\dagger d_{i,\sigma})
        +V\sum_{i,\sigma} 
       (d_{i,\sigma}^\dagger f_{i,\sigma}+f_{i,\sigma}^\dagger d_{i,\sigma})
       \nonumber \\
    & & \quad\quad +\epsilon_f\sum_{i,\sigma}n_{i,\sigma}^f
        +\frac{1}{2} U \sum_{i,\sigma}n_{i,\sigma}^fn_{i,-\sigma}^f
\label{eq:pam}
\end{eqnarray}
where the creation and destruction operators create and destroy
d-electrons on sites of a square lattice and f-electrons on localized
orbitals associated with these sites.
$n_{i,\sigma}^f=f_{i,\sigma}^\dagger f_{i,\sigma}$ is the number
operator for f-electrons.  Hopping only occurs between between
neighboring lattice sites (-$t$ term) and between a lattice site and its
orbital ($V$ term).

Because of the two bands, the model is half-filled when there are two
electrons per lattice site. At half-filling the model is said to be
symmetric when $\epsilon_f=-U/2$. For the symmetric model there is no
sign problem and standard QMC methods suggest that the model is an
insulating anti-ferromagnetic if $U>U_c\approx 2$. With the CPMC
method studying the magnetic properties of the doped model was
possible.

Upon doping with holes, the long-range anti-ferromagnetic order was
rapidly destroyed. Around three-quarters filling of the lower band, a
strong peak developed at the $(\pi,0)$ value of the spin structure
factor $S_\mathrm{ff}(k)$ for the $f$-electrons. This peak is shown in
Fig.~4, and its development is consistent with the resonance of two
degenerate spin-density waves with wave vectors $(\pi,0)$ and
$(0,\pi)$
\cite{bonca1}. Unestablished is whether this novel state is one of
long-range order.

\begin{figure}
\begin{center}
\epsfig{file=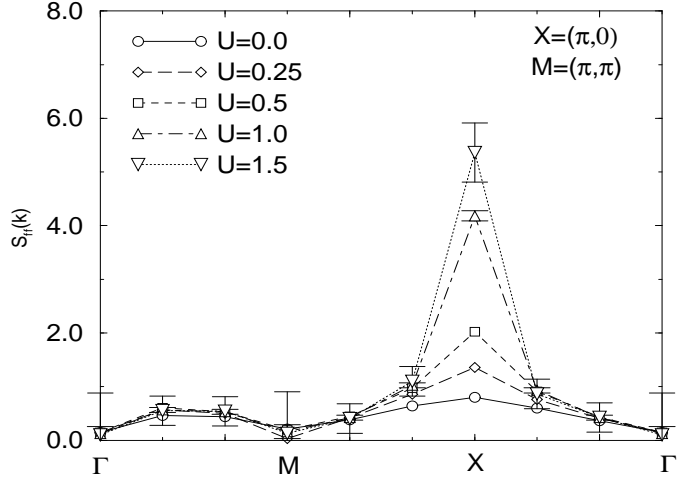,height=2.5in,width=3.5in}
\end{center}
\caption{Spin-spin correlation function $S_{\mathrm{ff}}({k})$ as a function of
wavenumber and $U$ for $\frac{3}{4}$-filling of the lower band
\cite{bonca1}. The plot in k-space is along symmetry lines in the
Brillouin zone}
\end{figure} 

\subsection{Three-Band Model}

The three-band model was constructed with the structure of
high-temperature superconductors in mind. The common structural
feature of these materials are $CuO_2$ planes. In addition, these materials
possess strong two-dimensional-like anisotropy. These properties have
focused attention on the physics in these planes as being the source
of the superconductivity.

The three-band model studied by the CPMC method represents the
Hamiltonian for the $CuO_2$ planes with only the most relevant $Cu$
and $O$ being kept
\begin{eqnarray}
H & = &            \sum_{<j,k> \sigma} t^{pp}_{jk} 
              (p^{\dagger}_{j \sigma}p_{k\sigma} +
               p^{\dagger}_{k\sigma}p_{j \sigma} )  + 
         \epsilon_p \sum_{j\sigma} n^{p}_{j\sigma} +
         U_p\sum_j n_{j\uparrow}^p n_{j\downarrow}^p  \nonumber   \\
 &    &   \hspace{2cm}      + \epsilon_d \sum_{i\sigma} n^{d}_{i\sigma} + 
            U_d\sum_i n_{i\uparrow}^d n_{i\downarrow}^d   \\
 &    &    + V^{pd}\sum_{<i,j>} n_{i}^d n_{j}^p \nonumber +
 \sum_{<i,j>\sigma} t^{pd}_{ij} (d^{\dagger}_{i\sigma}p_{j\sigma} +
	                      p^{\dagger}_{j\sigma}d_{i\sigma})   \nonumber 
\end{eqnarray}
In writing the Hamiltonian, we adopted the convention that the operator
$d^\dagger_{i,\sigma}$ creates a {\em hole\/} at a Cu $3d_{x^2 - y^2}$
orbital and $p^\dagger_{j,\sigma}$ creates a {\em hole\/} in an O $2p_x$ or
$2p_y$ orbital. $U_d$ and $U_p$ are the Coulomb repulsions at the Cu and O
sites respectively, $\epsilon_d$ and $\epsilon_p$ are the
corresponding orbital energies, and $V_{pd}$ is the nearest neighbor
Coulomb repulsion. As written, the model has a Cu-O hybridization $
t^{pd}_{ij} = \pm t_{pd} $ with the minus sign occurring $ j = i +
\hat{x}/2 $ and $j = i - \hat{y}/2 $ and also hybridization $
t^{pp}_{jk} = \pm t_{pp} $ between oxygen sites with the minus sign
occurring for $ k = j - \hat{x}/2 - \hat{y}/2 $ and $ k = j +
\hat{x}/2 + \hat{y}/2 $. The lattice is a planar $CuO_2$ structure
which has three atoms, one $Cu$ and two $O$, per unit cell. Hence the
non-interacting problem has three bands. In a bit of a switch,
convention for this model has half-filling corresponding to
half-filling of the anti-bonding (upper) band which is the lowest band in
the hole representation. When the Hamiltonian is
expressed in a hole representation, half-filling then corresponds to
one hole per unit cell.

The superconducting pairing correlation functions were studied with
the CPMC method, and the findings were similar to those found from the
studies on the Hubbard model: d-wave correlations are stronger than
the s-wave ones \cite{guerrero2}. For a given system size, increasing
$U$ suppress the pairing correlations. For a given $U$, increasing the
system size suppresses the pairing correlations.

For the three-band model the binding energy between two holes was also
calculated. This is a difficult calculation because it requires the
accurate calculation of the difference between nearly equal
energies. It is a significant calculation because the binding of holes
is a pre-requisite for superconductivity, phase separation, or stripe
formation.  

To calculate the binding energy for holes, we need to study the
half-filled case and then the 1 and 2 hole doped cases. In the systems
considered ($2 \times 2$, $4 \times 2$ and $6 \times 4$ {\em unit
cells\/}) the 2 hole doped case corresponds to a closed shell case.
The one hole case is one hole away from a closed shell, and the 
corresponding free-electron wave function is doubly degenerate. 
In this one-hole case the accuracy of the energy is as good as in the closed
shell case independently of the trial wavefunction used.
However, for the half-filled case, which is two holes away from a close 
shell, there are 4 degenerate free-electron states. 
If we used a trial state made from selecting
arbitrarily any one of the degenerate states or an arbitrary linear
combination of these states, the calculated energies would not be accurate
enough to compute the binding energy. Therefore, we used the
following procedure for the half-filled case: we
diagonalized the interacting part of the Hamiltonian in this
degenerate subspace, and obtained 2 states with energy proportional to
$U_d$ and 2 states with zero energy. Of the 2 states with zero energy
only one of them is a singlet. We used this state which is 
represented by a linear combination of two Slater 
determinants: $(c_{k_1,\uparrow}c_{k_1,\downarrow} -
c_{k_2,\uparrow}c_{k_2,\downarrow})|{\mathrm CS}\rangle/\sqrt{2}$.  In
Table 1, we compare the energies obtained using the CPMC with the one
and two Slater determinants trial wave function and energies obtained
from exact diagonalization. We see that using two Slater determinants
improves the accuracy by an order of magnitude or more. The accuracy
has become better than the closed shell case.

\begin{table}
\caption{
Comparison of the exact ground-state energy with the CPMC result with
a one and two Slater determinant trial wave function for a $2 \times
2$ system. Parameters are $\epsilon = \epsilon_p-\epsilon_d = 3$ and
$t_{pp} = U_p = 0$. The use of two Slater determinants as described in
the text improves the accuracy by an order of magnitude or more
\cite{guerrero2}.}
\begin{tabular}{ c  c  c c } \hline
\hspace{0.25cm} $U_d$\hspace{0.25cm} & \hspace{1cm}   CPMC 1SD \hspace{1cm}  & \hspace{1cm} CPMC 2SD \hspace{1cm}  &  \hspace{1cm} Exact\hspace{1cm} \\ \hline
1     &  -5.0613(3) &  -5.0764(2)  &   -5.076977   \\ 
2     &  -4.8475(9) &  -4.8789(7)  &   -4.880047   \\ 
4     &  -4.6073(9) &  -4.6615(6)  &   -4.661723   \\ 
6     &  -4.4884(9) &  -4.5468(6)  &   -4.547436   \\ \hline 
\end{tabular}
\end{table}

With this increased accuracy we found parameter ranges where holes
bind, values of the parameters where binding is optimal, and an
increase in binding energy with an increase in system size. Since the
appearance of hole binding seems decoupled from any enhancement of
superconducting correlation, an open question is the significance and
consequence of this binding.

\section{Parallelization}

If we are considering a system of $N_\mathrm{e}$ electrons and $N$
lattice sites, the CPU time scales as $N_\mathrm{w}N_\mathrm{e}N^2$
where $N_\mathrm{w}$ is the number of walkers. (The number of walkers
is typically of the order of $200$ to $1000$ with the larger number
usually need for the larger values of the interaction strengths.) This
number needs to be sufficiently large to insure an adequate
approximation to the ground state and is determined on the basis of
experience. The factor $N_\mathrm{e}N^2$ comes from the scaling of the
basic matrix operation that must be performed: The basic matrix
operations are matrix multiplication, inversion, re-orthogonalization,
and rank-one updates of matrix inverses. The dominant operation is the
matrix multiplication propagating a walker. The propagator is
represented by a $N\times N$ matrix, and each walker by two $N\times
N_{\sigma}$ matrices, i.e., one for each spin.

In general the method is CPU intensive as opposed to memory
intensive. As a consequence, the basic code usually fits in the memory
of one processor and the parallelization of the simulation can follow
one of several natural paths. The simplest path is to give each
processor a copy of the code, have it read different independent input
files, and compute independent runs. A less embarrassing
parallelization, and a run time reducing one, is to share the number of
walkers as equally as possible among as many processors as possible,
propagate the walkers on each processor independently, and combine the
results. The branched nature of the random walk however requires a
slightly different procedure.

Each walker carries a weight, and as the walker propagates, its weight
increases or decreases. Eventually the large weighted walkers
dominate, but propagating one of then cost as much computing time as
it does to propagate a small weighted walker that has little bearing
on the final results. Hence carrying the small weighted walkers
becomes inefficient. In these types of random walks, a standard
procedure is periodically eliminating walkers with small weights and
replacing each large weighted walker by many medium weighted walkers
whose total weight on the average is the same as the single
walker. There are several schemes to do this, and these procedures are
called population control. Population control prevents a single walker
from ultimately dominating the simulation and when used properly reduces the
variance of the computed results.

After a population control step, the loads on the processors are
unbalanced. The natural action then would be to redistribute the
load. Unless this load represents a relatively large number of walkers
per processor, a danger of introducing a bias into the simulation
exists by performing population control on a population too small to
be representative. Such a case is expected if several hundred walkers
distributed across a hundred or so processors.

Two other options for parallelization are easy to implement. One is to
use relatively few well-populated processors and have them
independently execute population control. (This is the procedure we
used for small and intermediate lattice sizes.) The other, which is
more effective in reducing run time for larger jobs when a large
number of processors are available, is doing population control by
moving all the walkers onto one processor, performing population
control there, and then uniformly re-distributing the
population. Because the amount of information passed between
processors is small and the amount of time needed for population
control is small, this procedure still achieves a nearly linear
reduction in computation time with the number of processors and is
simple to implement in any message passing environment.

\section{Concluding Remarks}

The constrained-path method has provided simulators of interacting
electron systems with a useful tool to study systems sizes impossible
by other means. With this method it has been possible to investigate
the timely and important question of the nature of superconducting
pairing correlations in candidate models for high temperature
superconductivity. The results obtained strongly suggest the absence
of long-range order in these models even though they do not provide a
rigorous proof of this absence.

While appearing similar to the fixed-node method long used for
continuum problem, it is not a fixed node method
\cite{carlson}. Appearing about the same time as the CPMC method was a
lattice version of the fixed-node method \cite{tenhaaf1,tenhaaf2}. The CPMC
method appears to give more accurate estimates of quantities like the
energy and correlation functions, but the fixed-node method can more
easily propagate some special trial states not expressible as a sum of
single Slater determinants.

The concept of constrained random walks has lead to two other quantum
Monte Carlo methods. One is a constrained random walk at finite
temperatures \cite{zhang4}. The initial results are promising. The
significance is that finite temperature methods have lacked effective
analogs even of the fixed-node method \cite{ortiz}. The other advance is the
constrained phase method. The method allows for the propagators and
Slater determinants to be complex valued: One has generalized the sign
problem to a phase problem. The ground-state wave functions of a system
of electrons in a magnetic field must be complex-valued. Making the
propagator complex-valued may be convenient if the system has
long-range forces.  Preliminary results on small systems in a magnetic
field are also promising.

The $16\times 16$ lattices size is most likely the largest for which
the simple parallelization scheme described above is convenient. Often
$N_\mathrm{e} \approx N$ so the computation scales roughly as
$N^3$. Thus simulating a $20\times 20$ system increases the run time
by a factor of 4 relative to the $16\times 16$ system. This increases
the simulation time from 2 weeks, for example, to 2 months. It is
becoming clear that one might need to simulate larger lattice sizes to
insure finite-size effects are not influencing the results. Going to
these larger systems will require distributing the matrix operations
across many processors. Such operations are all fundamental, and
procedures for doing this type of parallelization exist. We will be
exploring their utilization in the near future.


\begin{thebibliography}{9}

\bibitem{zhang1}
Shiwei Zhang, J. Carlson, and J. E. Gubernatis,
Constrained path Monte Carlo method for fermion ground states,
Phys. Rev. Lett. {\bf 74} (1995) 3652-3656.

\bibitem{zhang2}
Shiwei Zhang, J. Carlson, and J. E. Gubernatis,
Constrained path Monte Carlo method for fermion ground states,
Phys. Rev. B {\bf 55} (1997) 7464-7477.

\bibitem{loh} For example see, E. Y. Loh, Jr., and J. E. Gubernatis,
in: W. Hanke and Yu. V. Kopaev, eds., 
{\em Electronic phase transitions\/} (North-Holland, Amsterdam, 1992)
177-235.

\bibitem{reynolds}
For example, P. J. Reynolds, D. M. Ceperley, B. J., Alder, and
W. A. Lester, Jr.,
Fixed-node Monte Carlo for molecules,
J. Chem. Phys. {\bf 92} (1982) 5593-5603.

\bibitem{bonca2}
J. Bon\v ca and J. E. Gubernatis,
Quantum Monte Carlo by message passing,
in: D. A. Browne, J. Calloway, J. P. Draayer, R. W. Haymaker,
R. K. Kalia, J. E. Tohline, and P. Vashishta,
{\em High performance computing and its applications in the physical
sciences \/}
(World Scientific, Singapore, 1994) 60-71.

\bibitem{guerrero1}
M. Guerrero, G. Ortiz, and J. E. Gubernatis,
Correlated wave functions and the absence of long-range order in
numerical studies of the Hubbard model,
Phys. Rev. B {\bf 59} (1999) 1706-1711.
 
\bibitem{schmidt}
K. E. Schmidt and S. Fantoni,
Phys. Lett. B {\bf 446} (1999) 99-103.

\bibitem{zhang3}
Shiwei Zhang, J. Carlson, and J. E. Gubernatis,
Pairing Correlations in the two-dimensional Hubbard model,
Phys. Rev. Lett. {\bf 78} (1997) 4486-4490.

\bibitem{huang} Z. Huang, H. Q. Lin, and J. E. Gubernatis,
unpublished.

\bibitem{zhang4} Shiwei Zhang, unpublished.

\bibitem{mazumdar} S. Mazumdar, S. Ramasesha, R. T. Clay, and
D. K. Campbell,
Phys. Rev. Lett. {\bf 82} 1522 (199).

\bibitem{bonca1}
J. Bon\v ca and J. E. Gubernatis,
Effects of pairing correlations on spin correlations in the periodic
Anderson lattice,
Phys. Rev. B {\bf 58} (1998) 6992-7001.

\bibitem{guerrero2}
M. Guerrero, J. E. Gubernatis, and Shiwei Zhang,
Quantum Monte Carlo study of hole binding and pairing correlations in the
three-band Hubbard model,
Phys. Rev. B {\bf 57}, (1998) 11980-11988.

\bibitem{carlson}
J. Carlson, J. E. Gubernatis, G. Ortiz, and Shiwei Zhang,
Phys.  Rev. B {\bf 59} 12788 (1999) 12788-12798.

\bibitem{tenhaaf1}
D. F. B. ten Haaf, H. J. M. van Bemmel, J. M. J. van Leeuwen, W. van
Saarlos, and G. An,
Fixed node quantum Monte Carlo method for lattice fermions,
Phys. Rev. Lett. {\bf 72} (1994) 2442-2446.

\bibitem{tenhaaf2}
D. F. B. ten Haaf, H. J. M. van Bemmel, J. M. J. van Leeuwen, W. van
Saarlos, and D. M. Ceperley,
Proof for an upper bound in fixed-node Monte Carlo for lattice fermions,
Phys. Rev. B {\bf 51} (1995) 13039-13045.

\bibitem{ortiz}
G. Ortiz, J. Carlson, and J. E. Gubernatis,
unpublished.

\end{thebibliography}
\end{document}